\RequirePackage{fix-cm}
 \documentclass[twocolumn]{svjour3}          
\smartqed  

\usepackage{nicefrac}
\usepackage{color}
\usepackage{amsmath}
\usepackage{amssymb}
\usepackage{graphicx}
\newcommand{\figWidth}{0.95\columnwidth}

\newcommand{\eqRef}[1]{Eq.~(\ref{#1})}
\newcommand{\figRef}[1]{Fig.~\ref{#1}}

\newcommand{\tabRef}[1]{Tab.~\ref{#1}}

\begin{document}

\title{Oblique Impact of Frictionless Spheres}
\subtitle{On the Limitations of Hard Sphere Models for Granular Dynamics}
\author{Patric M\"uller \and Thorsten P\"oschel
}                     
\institute{Patric M\"uller \and Thorsten P\"oschel\at
				Institute for Multiscale Simulation\\
Universit\"at Erlangen-N\"urnberg\\
N\"agelsbachstra\ss{}e 49b\\
91052 Erlangen\\
Germany\\
\email{patric.mueller@cbi.uni-erlangen.de}           
}

\date{Received: \today / Revised version: date}
\maketitle
\abstract{When granular systems are modeled by frictionless hard spheres, particle-particle collisions are considered as instantaneous events. This implies that while the velocities change according to the collision rule, the positions of the particles are the same before and after such an event. We show that depending on the material and system parameters, this assumption may fail. For the case of viscoelastic particles we present a universal condition which allows to assess whether the hard-sphere modeling and, thus, event-driven Molecular Dynamics simulations are justified. 
\keywords{Granular Gases \and Hard Sphere Model \and Coefficient of Normal Restitution
\and Viscoelastic Spheres \and event-driven Molecular Dynamics 
}
\PACS{
      {45.50.Tn }{Collisions}   \and
      {45.70.-n }{Granular systems}
     } 
} 
\section{Introduction}
\label{sec:intro}
Hard sphere modeling of granular systems assumes that the dynamics of the system may be described as a sequence of {\em instantaneous} events of {\em binary} collisions. In between the collisions the particles move freely along straight lines, or ballistic trajectories in presence of external fields like gravity. The hard-sphere model of particle collisions is the foundation of both Kinetic Theory of granular matter, based on the Boltzmann equation, e.g. \cite{goldhirsch2003,GranularGases,PROCEEDINGS}, and event-driven Molecular Dynamics (eMD) of granular matter, e.g. \cite{lubachevsky1991,rapaport1980,poeschel2005}. 

In hard sphere approximation, the inelastic collision of frictionless spheres $i$ and $j$ located at $\vec{r}_i$ and $\vec{r}_j$ traveling at velocities $\dot{\vec{r}}_i$ and $\dot{\vec{r}}_j$ is, thus, characterized by the collision rule describing the instantaneous exchange of momentum between the colliders,
\begin{equation}
\label{eq:epsNCorrectDef}
 \left(\dot{\vec{r}}_i^{\prime}-\dot{\vec{r}}_j^\prime\right)\cdot\hat{e}^{\:\prime}_{r}
=-\varepsilon\left(\dot{\vec{r}}^0_i-\dot{\vec{r}}_j^0\right)
\cdot\hat{e}_{r}^0\,.
\end{equation}
with the unit vector $\hat{e}_{r}\equiv \left(\vec{r}_i-\vec{r}_j\right)/\left|\vec{r}_i-\vec{r}_j\right|$. Upper index $0$ describes values just before the collision, primed values describe postcollisional values. The inelasticity is characterized by the coefficient of (normal) restitution $\varepsilon$.

The instantaneous character of the collisions implies that as the result of a collision only the velocities of the particles change but not their positions, $\vec{r}_i^\prime=\vec{r}_i^0$, $\vec{r}_j^\prime=\vec{r}_j^0$ and, thus, $\hat{e}^{\prime}_{r}\equiv\hat{e}_{r}^0$. With this, \eqRef{eq:epsNCorrectDef} turns into
\begin{equation}
\label{eq:epsNCorrectDefHS}
 \left(\dot{\vec{r}}_i^{\prime}-\dot{\vec{r}}_j^\prime\right)\cdot\hat{e}^0_{r}
=-\varepsilon^{\text{HS}}\left(\dot{\vec{r}}^0_i-\dot{\vec{r}}_j^0\right)
\cdot\hat{e}_{r}^0\,
\end{equation}
which allows to compute the postcollisional velocities successively for all collisions in the system, which is the basic idea of event-driven Molecular Dynamics (EMD). Provided the system may be described as hard spheres undergoing instantaneous collisions, EMD may be by orders of magnitude more efficient than ordinary MD integrating Newton's equation of motion. Recently, extremely efficient algorithms for EMD simulations have been developed, e.g. \cite{Bannerman}.

As physical particles are not perfectly hard but the collision is governed by {\em finite} interaction forces, the hard sphere model is an idealization whose justification may be challenged. Especially in view of its importance for Kinetic Theory and numerical simulation techniques. In particular, for finite duration of the collisions the unit vector $\hat{e}_r$ may rotate during a collision by the angle 
\begin{equation}
 \label{eq:alphaDef}
 \alpha\equiv\arccos\left(\hat{e}_{r}^0\cdot\hat{e}_{r}^{\:\prime}\right)
\end{equation}
invalidating Eq. \eqref{eq:epsNCorrectDefHS} and, therefore, the hard-sphere approximation. While this angle is negligible for approximately central impacts of relatively stiff spheres, it is not for oblique impacts of soft spheres \cite{becker2008}.

Within this work we quantify under which conditions and to what extend the condition, $\hat{e}_{r}^\prime\approx\hat{e}^0_r$, of the hard sphere assumption fails. Aim of the present paper is to provide a universal condition which allows for {\em arbitrary} collisions of {\em arbitrary} elastic spheres to assess whether the hard sphere model is acceptable for the description of particle collisions. Thus, we discriminate wether the hard sphere model is acceptable for systems, characterized by (i) a set of material parameters, (ii) particle sizes and (iii) a typical (thermal) impact velocity. To generalize our result to the case of inelastic collisions, we show that regarding the rotation angle $\alpha$, elastic spheres are the marginal case, that is, if $\hat{e}_{r}^\prime\approx\hat{e}^0_r$ holds true for elastic particles, it \emph{certainly} holds true for inelastic particles.

\section{Collision of spheres}
\label{sec:smoothSpheres}
Consider two colliding spheres with the masses $m_i$ and $m_j$ located at $\vec{r}_i$ and $\vec{r}_j$ and traveling with velocities $\dot{\vec{r}}_i$ and $\dot{\vec{r}}_j$.
With the interaction force $\vec{F}$, their motion is described by 
\begin{equation}
\label{eq:newton}
m_{\text{eff}}\ddot{\vec{r}}=\vec{F}\,,~~~~~
M\ddot{\vec{R}}=\vec{0}
\end{equation}
where 
\begin{equation}
\label{eq:COMAndRelCoordDef}
 \vec{R}\equiv \frac{m_i\vec{r}_i+m_j\vec{r}_j}{m_i+m_j}\,,~~~\vec{r}=\vec{r}_i-\vec{r}_j\,,~~~m_{\text{eff}}=\frac{m_im_j}{m_i+m_j}
\end{equation}
are the center of mass coordinate, the relative coordinate and the effective mass, respectively. The center of mass moves due to external forces such as gravity and separates from the relative motion which in turn contains the entire collision dynamics. 

For frictionless particles, the interaction force exclusively acts in the direction of the inter-center unit vector, $\vec{F}=F_n \hat{e}_r$, that is,
there is no tangential force and, thus, the particles' rotation is not affected by the collision. During the collision the (orbital) angular momentum is conserved which allows for the definition of a constant unit vector:
\begin{equation}
\label{eq:angMomDef}
 \vec{L}= m_{\text{eff}}\:\vec{r}\times\dot{\vec{r}}\equiv L\hat{e}_L\,.
\end{equation}
Thus, with the coordinate system spanned by
\begin{equation}
\hat{e}_{x}\equiv\hat{e}_r^0\,,~~~~
\hat{e}_{z}\equiv\hat{e}_L\,,~~~~
\hat{e}_{y}\equiv\hat{e}_z\times\hat{e}_x\,,
\end{equation}
and with its origin in the center of mass $\vec{R}$, the collision takes place in the $\hat{e}_{x}$-$\hat{e}_{y}$--plane.\footnote{For central collisions we have $\vec{L}=\vec{0}$. In this case $\hat{e}_z$ may be any unit vector perpendicular to $\hat{e}_x$, ($\hat{e}_x\cdot\hat{e}_z=0$).} 
\begin{figure}[h!]
 \includegraphics[width=\figWidth]{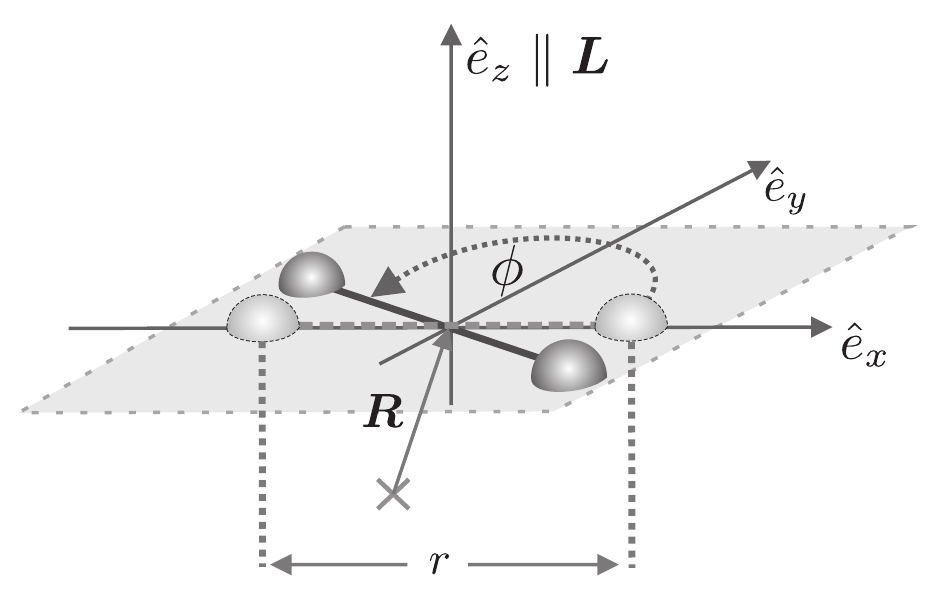}
 \caption{Illustration of the used polar coordinates (see text)}
 \label{fig:polarDef}
\end{figure}
In the collision plane we formulate the equation of motion in polar coordinates
$\left\{r,\varphi\right\}$ (see \figRef{fig:polarDef}):
\begin{equation}
\label{eq:newtonPolar}
 m_{\text{eff}}r^2\dot{\varphi}=L\,,~~~~
 m_{\text{eff}}\ddot{r}=F_c+F_n=m_{\text{eff}}r\dot{\varphi}^2+F_n\,,
\end{equation}
with the centrifugal force $F_c$. Together with the inital conditions 
\begin{equation}
 r(0)=r^0\,,~~~~
 \dot{r}(0)=\dot{r}^0\,,~~~~
 \varphi(0)=0\,,
\end{equation}
\eqRef{eq:newtonPolar} fully describes the collision dynamics for an arbitrary
normal force $F_n$. The collision terminates at time $t=\tau$ where \cite{schwager2007,schwager2008}
\begin{equation}
\label{eq:tauDef}
 \dot{r}(\tau)>0\text{~~~and~~~}F_n=0.
\end{equation}
Inserting the first equation of \eqRef{eq:newtonPolar} into the second, we obtain 
\begin{equation}
 m_{\text{eff}}\ddot{r}=\frac{L^2}{m_{\text{eff}}r^3}+F_n
\end{equation}
which fully governs the radial dynamics of the problem.

Note that in contrast to earlier work \cite{schwager2007,schwager2008}  where the coefficient of normal restitution was derived from force laws $F_n$ here we allow the normal vector $\hat{e}_r$ to rotate during the collision and do not neglect the resulting centrifugal force.

Since for {\em any} finite interaction forces, the duration $\tau$ of a collision is finite, for non-central collisions, $\vec{L}\neq0$, during the collision the spheres rotate around their center of mass, that is, $\hat{e}_r^{\:\prime}\neq\hat{e}^0_r$ and $\alpha\neq0$, see \eqRef{eq:alphaDef}.

It is frequently stated that the hard sphere approximation and thus event-driven simulations are always justified for dilute systems where the mean free flight time of the particles is large compared to the typical collision time. Obviously, this condition is insufficient. It may be shown that the characteristics of dilute granular gases such as the coefficient of self diffusion sensitively depends on the rotation of the unit vector \cite{future}.

\section{Elastic Spheres}
\label{sec:Elastic}
\subsection{Dimensonless Equation of Motion}
\label{sec:Dimensionless}
The collision of elastic spheres obeys Hertz' contact force \cite{hertz1881},
\begin{equation}
\label{eq:fHertz}
 F_n=F_n^{\text{el}}=\rho(l-r)^{3/2}\,,~~~~l\equiv r^0=R_i+R_j\,,
\end{equation}
where $l$ denotes the distance between the particle centers at the moment of impact. The quantity $\xi\equiv l-r$ is often referred to as the deformation or mutual compression. The elastic constant $\rho$ reads
\begin{equation}
  \label{eq:rhodef}
 \rho\equiv\frac{2Y\sqrt{R_{\text{eff}}}}{3(1-\nu^2)}
\end{equation}
where $Y$, $\nu$ and $R_{\text{eff}}$ stand for the Young modulus, the Poisson ratio and the effective radius $R_{\text{eff}}=R_iR_j/(R_i+R_j)$, respectively.

Writing the general equation of motion \eqRef{eq:newtonPolar} with the force given by Eq. \eqref{eq:fHertz} and measuring length in units of $X$ and time in units of $T$ \cite{schwager2008}, 
\begin{equation}
  \label{eq:XTdef}
  X\equiv\frac{(-\dot{r}^0)^{4/5}}{k^{2/5}}\,,~~~T\equiv\frac{1}{k^{2/5}(-\dot{r}^0)^{1/5}}\,,~~~ k\equiv\frac{\rho}{m_{\text{eff}}}
\end{equation}
we obtain
\begin{eqnarray}
\label{eq:newtonPolarScaledA}
\frac{\mathrm d \varphi}{\mathrm d
\tilde{t}}&=&\frac{c_\varphi}{\tilde{r}^2}\,,~~~~~\\
\label{eq:newtonPolarScaledB}
 \frac{\mathrm d^2\tilde{r}}{\mathrm d\tilde{t}^2}&=&
 \tilde{r}\left(\frac{\mathrm d \varphi}{\mathrm d \tilde{t}}\right)^2+
 \left(\tilde{l}-\tilde{r}\right)^{3/2}
\end{eqnarray}
with
\begin{equation}
\label{eq:scalingDef}
 \tilde{t}\equiv\frac{t}{T},\quad\tilde{r}\equiv\frac{r}{X},\quad\tilde{l}
\equiv\frac{l}{X}
\end{equation}
and
\begin{equation}
\label{eq:cPhiCDisDef}
 c_\varphi\equiv\frac{T}{X^2}\frac{L}{m_{\text{eff}}}\,.
\end{equation}
The scaled initial conditions read
\begin{equation}
\label{eq:initCondScaled}
 \varphi(0)=0,\quad\tilde{r}(0)=\tilde{l}\quad\text{and}\quad\dot{\tilde{r}}(0)=1.
\end{equation}
According to Eqs. \eqref{eq:newtonPolarScaledA} and \eqref{eq:newtonPolarScaledB} together with the initial conditions  \eqRef{eq:initCondScaled}, the binary collision of frictionless elastic spheres is described by only two free parameters: $\tilde{l}$ and $c_\varphi$.

\subsection{Rotation of the Normal Vector}
\label{sec:Rotation}
\begin{figure}[h!]
 \includegraphics[width=\figWidth,viewport=79 225 445 366,clip]{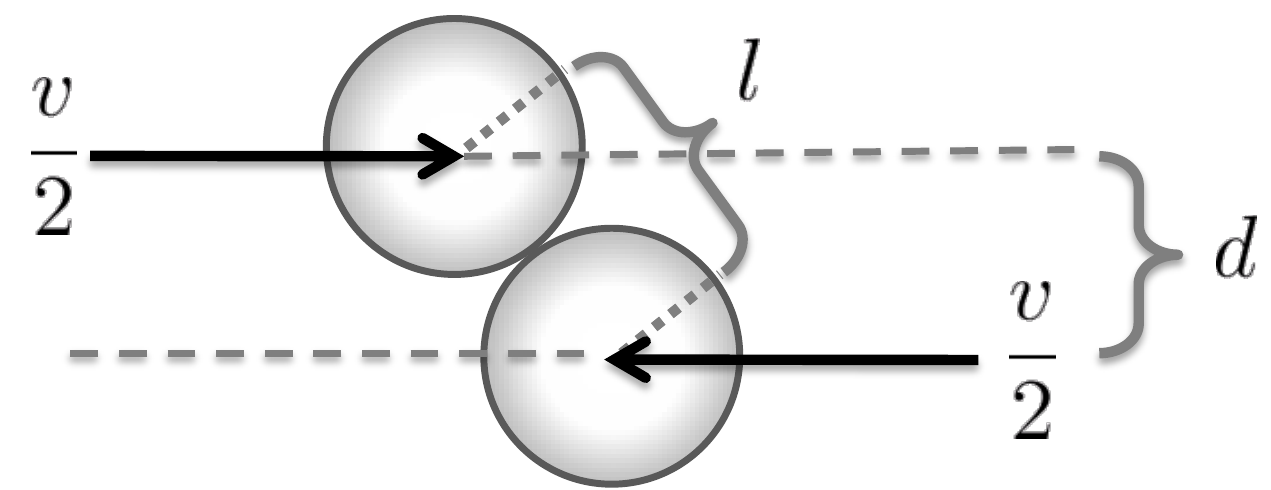}
 \caption{Eccentric binary collision of spheres. The sketched situation corresponds to the eccentricity $d/l\approx0.8$ where the rotation angle drawn in Fig. \ref{fig:alphaVsDOverD} adopts its maximum.}
 \label{fig:excCollSetup}
\end{figure}

We solve the equations of motion \eqref{eq:newtonPolarScaledA}, \eqref{eq:newtonPolarScaledB} and \eqref{eq:initCondScaled} to obtain the rotation $\alpha$ of the unit vector, $\hat{e}_r$, given by Eq. \eqref{eq:alphaDef} as a consequence of the collision of elastic spheres. This rotation occurs for oblique impacts and is commonly neglected in hard sphere simulations as well as in the Kinetic Theory of granular gases. Obviously, $\alpha$ depends on the material properties, the particle sizes and on the geometry of the collision as sketched in Fig. \ref{fig:excCollSetup}. The condition $\alpha\approx0$ will be interpreted as an index for the justification of the hard-sphere approximation for a given system.

\begin{figure}[h!]
 \includegraphics[width=\figWidth]{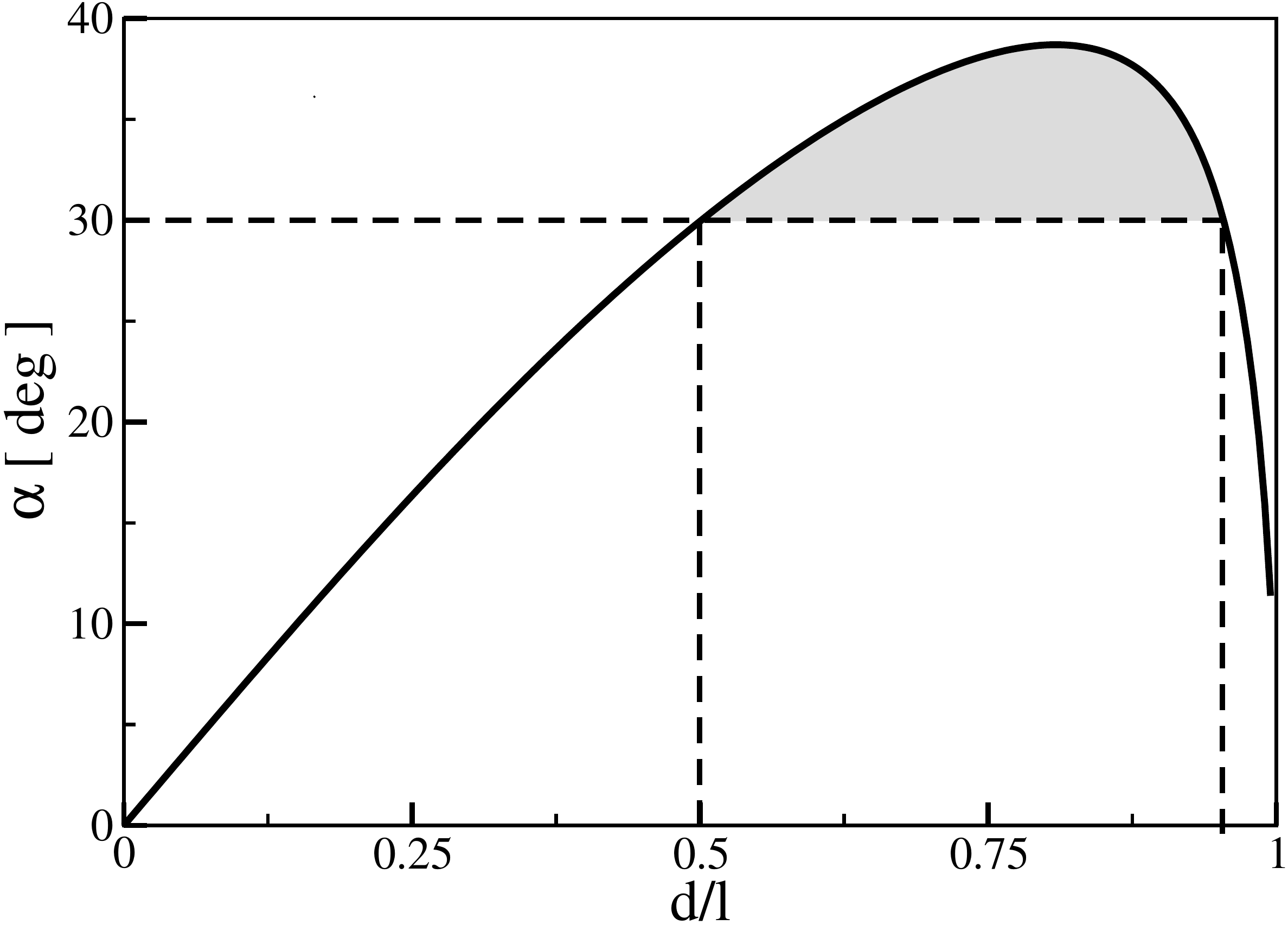}
\caption{Rotation angle $\alpha$ of the unit vector $\hat{e}_r$ as a function of the impact eccentricity $d/l$ (see \figRef{fig:excCollSetup}) for rubber spheres, parameters specified in the text. The marked area shows the interval where $\alpha > 30^\circ$ corresponding to $65\%$ of all collisions when molecular chaos is assumed.}
\label{fig:alphaVsDOverD}
\end{figure}

To illustrate the fact that the rotation of the unit vector is by far not a small effect even for rather common systems, in  
\figRef{fig:alphaVsDOverD} we plot the angle $\alpha$ as a function of the impact eccentricity $d/l$ (see  \figRef{fig:excCollSetup}). The system parameters (in physical units) are: radii $R_1=R_2=$0.1\,m, material density $\rho_m=1140$\,kg$/$m$^3$, Young modulus $Y=10^7$\,N$/$m$^2$, Poisson ratio $\nu=0.4$, impact velocity $20$\,m$/$s. 

As expected, the rotation vanishes for central collisions. The rotation adopts its maximum for $d/l\approx 0.8$ (this situation corresponds to the sketch in \figRef{fig:excCollSetup}) where it can easily reach values of $\alpha\approx 40^o$. The position of the maximum may surprise since in the Kinetic Theory it is frequently assumed that if at all only rare {\em glancing collisions} might deserve a special consideration. Assuming molecular chaos, that is, $e\equiv d/l$ is distributed as $\mathrm dp(e)=2e\:\mathrm de$, and the parameters given above, about $65\%$ of the collisions lead to a rotation angle $\alpha > 30^\circ$ (marked interval in \figRef{fig:alphaVsDOverD}). Consequently, the rotation of the unit vector $\hat{e}_r$ is a very significant effect for granular gases.

\subsection{Universal Description of the Rotation Angle}
For elastic particles the dimensionless equation of motion of the collision (Eqs. \eqref{eq:newtonPolarScaledA}, \eqref{eq:newtonPolarScaledB} and \eqref{eq:initCondScaled}) is fully specified by two independent parameters, $\tilde{l}$ and $c_\varphi$, defined in Eqs. \eqref{eq:scalingDef} and \eqref{eq:cPhiCDisDef}. Therefore all material and system parameters may be mapped to a point in the $(\tilde{l}\,, c_\varphi)$-space.

The rotation angle $\alpha$ can be determined by the following procedure: 
\begin{enumerate}
\item Determine the dimensionless parameters: \\$\{Y,\nu,R,\rho_m,v,d/l\}$ $\to$ $\{\tilde{l}, c_\varphi\}$

\item Solve numerically the equations of motion,\\ Eqs. (\ref{eq:newtonPolarScaledA},\ref{eq:newtonPolarScaledB},\ref{eq:initCondScaled}) for $0\le t\le \tau$ where $\tau$ is the time when the collision terminates. $\tau$ is determined by the conditions $\ddot{\tilde{r}}(\tau)=0$ and $\dot{\tilde{r}}(\tau)>0$ (see Eq. \eqref{eq:tauDef}).

\item The rotation angle is obtained from $\alpha=\varphi(\tau)$.

\end{enumerate}

\begin{figure}
 \includegraphics[width=\figWidth]{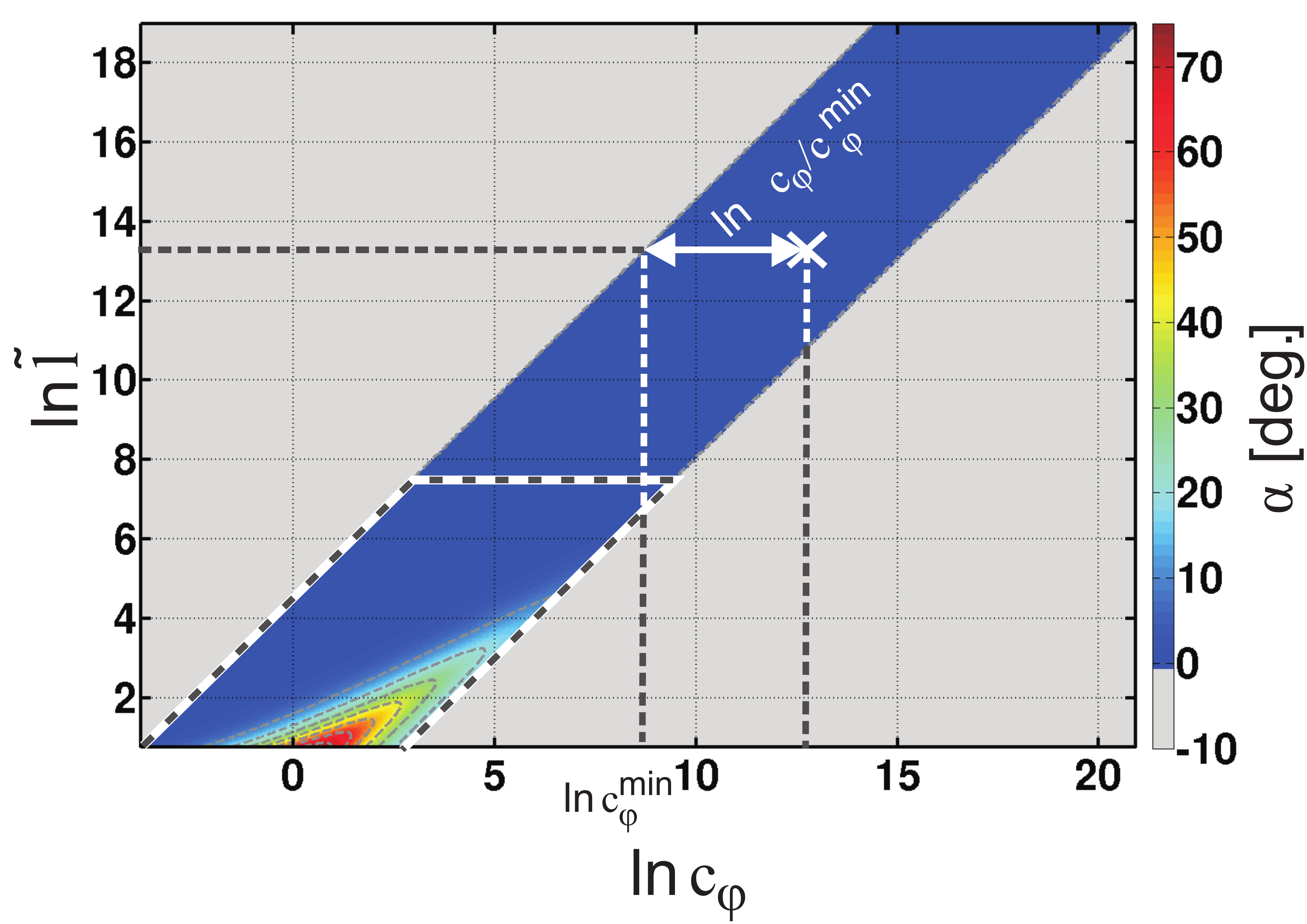}
\caption{Rotation angle $\alpha$ as function of $\tilde{l}$ and $c_\varphi$.
Grey regions indicate points which do not correspond to any combination of
parameters given in Tab. \ref{tab:physParamSpace}. Only combinations inside the
dashed region may lead to a noticeable rotation angle.}
\label{fig:angPColor}
\end{figure}

We performed this procedure for a wide range of relevant (physical) parameters given in Table \ref{tab:physParamSpace}. 
\begin{table}[h!]
\centering
\begin{tabular}{lllll}
\hline\hline\\
&unit&min.&max.&\\
$Y$&[$10^9$ N$/$m$^2$]&$0.01$&$100$&Young's modulus\\
$\nu$&-&$0.2$&$0.5$&Poisson ratio\\
$R$&[m]&$0.001$&$0.1$&particle radius\\
$\rho_m$&[kg$/$m$^3$]&$250$&$3250$&material density\\
$v$&[m$/$s]&$0.001$&$25$&impact velocity\\
$d/l$&-&$0.01$&$0.99$&eccentricity\\\hline
\end{tabular}
\caption{Physical parameter space scanned to obtain \figRef{fig:angPColor}. For the definition of the impact velocity and the eccentricity we refer to \figRef{fig:excCollSetup}.}
\label{tab:physParamSpace}
\end{table}
In dimensionless variables, this range corresponds to the interval
\begin{equation}
2.12 \le \tilde{l}\le 1.8\cdot10^9\,,~~~
2.12\cdot10^{-2} \le c_\varphi\le 1.26\cdot10^9\,.
\end{equation} 
Figure \ref{fig:angPColor} shows the rotation angle $\alpha$ as a function of $\tilde{l}$ and $c_\varphi$ on a double logarithmic scale.

Using the definitions of $\tilde{l}$, $c_\varphi$ and $\vec{L}$, Eqs. (\ref{eq:scalingDef},\ref{eq:cPhiCDisDef},\ref{eq:angMomDef}) and 
\begin{equation}
  \label{eq:2}
  \frac{X}{T}=-\dot{\vec{r}}^0=v\sqrt{1-\left(\frac{d}{l}\right)^2}
\end{equation}
which follows from the definitions, Eq. \eqref{eq:XTdef}, and geometry (see \figRef{fig:excCollSetup}), one obtains
\begin{equation}
\label{eq:lTildeOfCPhiDOverD}
 \ln\tilde{l}=\ln\left(c_\varphi\right) + \frac{1}{2}\ln\left[\left(\frac{l}{d}\right)^2-1\right]\,.
\end{equation}
This equation provides some insight into the structure of \figRef{fig:angPColor} and allows for a more intuitive presentation of the result. For fixed eccentricity $d/l$ due to Eq. \eqref{eq:lTildeOfCPhiDOverD}, in the double logarithmic scale used in \figRef{fig:angPColor}, $\tilde{l}$ is a linear function of $c_\varphi$ with slope 1. That is, all collisions taking place at the same impact eccentricity $d/l$ are located on a straight line of slope 1 in the $(\ln c_\varphi,\ln\tilde{l})$-space. The position along this line is then determined by the remaining system parameters. 

The chosen interval, $0.01\le d/l\le0.99$, see \tabRef{tab:physParamSpace}, implies that the intercept, of all possible straight lines \ given by \eqRef{eq:lTildeOfCPhiDOverD} is bound to the range 
\begin{equation}
  \label{eq:3}
 -1.95 \le \left.\ln\tilde{l}\right|_{c_\varphi=0} \le 4.61\,,
\end{equation}
which explains the stripe structure of the data in \figRef{fig:angPColor}. All $(\ln\tilde{l},\ln c_\varphi)$-pairs outside the colored stripe {\em cannot} be adopted for any combination of the parameters listed in Tab. \ref{tab:physParamSpace} which is indicated by the gray areas in \figRef{fig:angPColor}.

\begin{figure}[h!]
 \includegraphics[width=\figWidth]{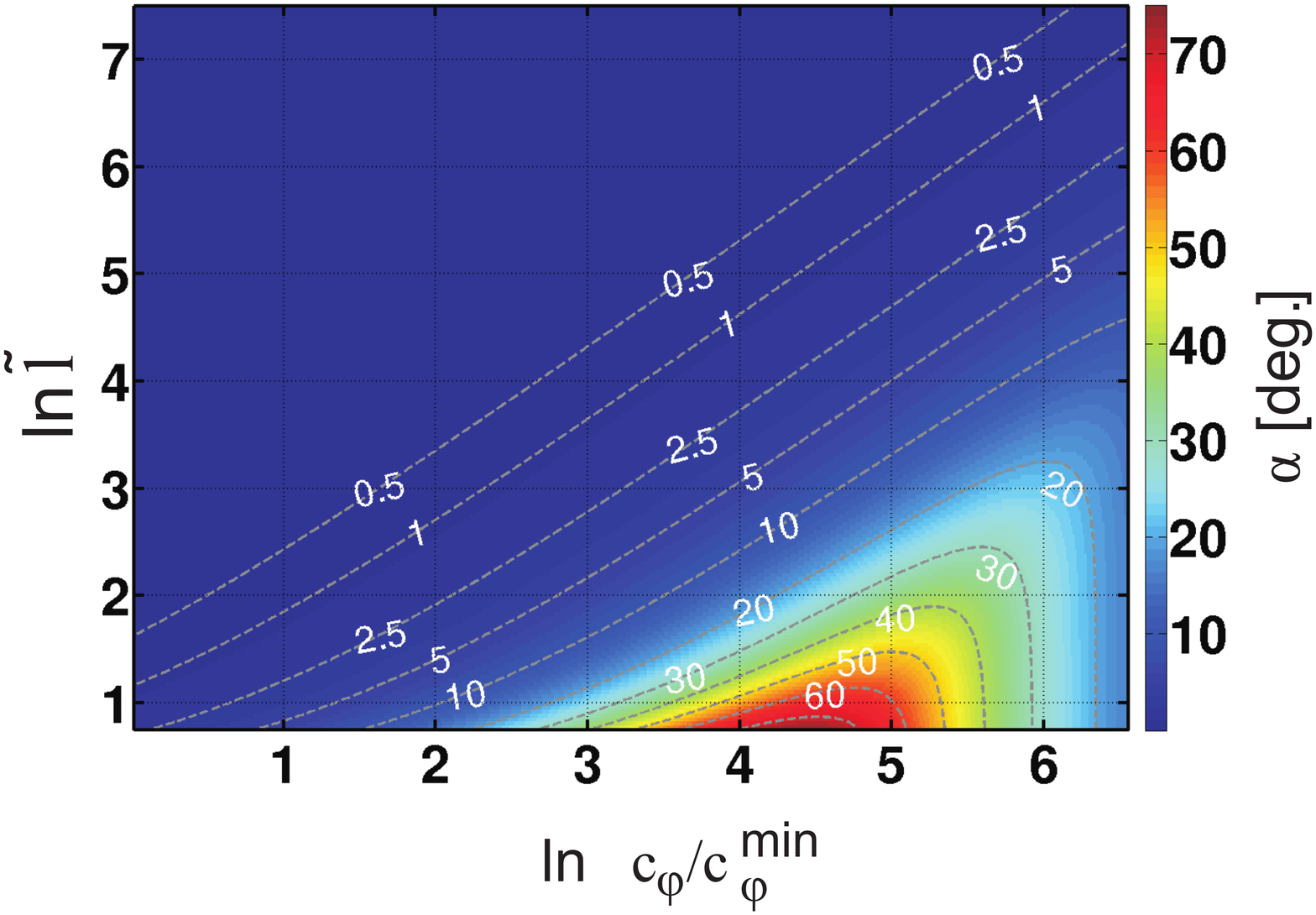}
\caption{Rotation angle $\alpha$ as function of $\tilde{l}$ and $c_\varphi/c_\varphi^{\min}$. The figure shows the region of the parameter space where $\alpha$ adopts noticeable values (dashed region in \figRef{fig:angPColor}). Additionally several isolines of constant rotation angle $\alpha$ are shown.}
\label{fig:angPColorCutout}
\end{figure}

Figure~\ref{fig:angPColor} indicates that among all studied combinations of parameters only those for $-3\lessapprox\ln c_\varphi\lessapprox9$ and $1\lessapprox\ln\tilde{l}\lessapprox8$ (dashed region in \figRef{fig:angPColor}) may lead to a noticeable rotation angle $\alpha$ or a significant deviation from the hard sphere model respectively. Therefore, we study this region with a higher resolution of $c_\varphi$ and $\tilde{l}$, see \figRef{fig:angPColorCutout}. In order to avoid drawing the irrelevant gray regions, we plot the data over $\ln c_\varphi-\ln c_\varphi^{\min}$ instead of $\ln c_\varphi$ with 
\begin{equation}
\ln c_\varphi^{\min} = 
\ln \tilde{l} - \frac12\ln\left[ \frac{1}{\left(\frac{d}{l}\right)^2_{\min}}-1\right]\approx \ln\tilde{l}-4.61\,.
\end{equation}
as obtained from Eq. \eqref{eq:lTildeOfCPhiDOverD} with $(d/l)_{\min}$ taken from Tab. \ref{tab:physParamSpace} (see illustration of $\ln c_\varphi-\ln c_\varphi^{\min}$ in \figRef{fig:angPColor}).

The isolines of constant rotation angle $\alpha$ drawn in \figRef{fig:angPColorCutout} indicate that there is a rather sharp transition between the regions where $\alpha\approx0$ and $\alpha\gg0$ in the $(\ln c_\varphi,\ln\tilde{l})$-space. Hence, regarding the rotation of the unit vector $\hat{e}_r$, the regions in the parameter space where the hard sphere model is a justifiable approximation are clearly separated from those, where the hard sphere approximation is questionable.

\subsection{Confidence Regions of the Hard Sphere Model}
For practical applications one might wish to know whether a given set of material and system parameters allows for a hard-sphere description. Besides other criteria, the maximum rotation angle which can be expected for these parameters, is an important criterion. For the following we assume that the rotation angle $\alpha_{c}$ is marginally acceptable for the hard sphere approximation and provide a simple approximate method do decide whether the given system fulfills this criterion.  

\figRef{fig:angPColorCutout} shows that for rotation angles up to about $15^\circ$, the isolines of constant rotation angle are approximately straight lines of slope $\overline{m}\approx0.84$ on average. The corresponding intercept $t_{\alpha_c}$ increases with the isoline value $\alpha_c$. From \figRef{fig:angPColorCutout} we obtain $t_{1^\circ}\approx0.9$, $t_{5^\circ}\approx -0.29$, $t_{10^\circ}\approx -0.74$ and $t_{15^\circ}\approx -0.85$. 

We specify a collision by
\begin{equation}
  \begin{split}
\ln\:\frac{c_\varphi}{c_\varphi^\text{min}}&=4.61-\frac{1}{2}\ln\left[\left(\frac{l}{d}\right)^2-1\right],\\
\tilde{l} &=\left[\frac{2Y\sqrt{R_{\text{eff}}}}{3(1-v^2)m_{\text{eff}}}\right]^{2/5}\left[v\sqrt{1-(d/l)^2}\right]^{-4/5}l    
  \end{split}
\end{equation}
and define
\begin{equation}
 D_{\alpha_c}\equiv \ln \tilde{l}-\left(\overline{m} \ln\:(c_\varphi/c_\varphi^\text{min})+t_{\alpha_c}\right)\,.
\end{equation}
$D_{\gamma}>0$ indicates that the maximally expected rotation angle is smaller than $\alpha_c$, that is, the hard sphere model is acceptable for this situation.   

\section{Inelastic Spheres}
\label{sec:viscelSpheres}
\subsection{Equation of Motion}
The main conclusion of this Section will be that inelastic interaction forces which are, perhaps, the most characteristic feature of granular materials, do not lead to an increase of the rotation angle $\alpha$ as compared with the elastic case detailed in the previous Section. Here, we exemplary discuss a particular dissipation mechanism, the viscoelastic model which is widely used for modeling granular systems, e.g. \cite{kruggelEmden2007,stevens2005,schaefer1996}. Many other dissipative interaction forces as plastic deformation, linear dashpot damping, etc. lead to very similar results.

The collision of viscoelastic spheres is characterized by the interaction force \cite{brilliantov1996}
\begin{equation}
\label{eq:fNViscel}
F_n=F_n^{\text{el}}+ F_n^{\text{dis}} =\rho(l-r)^{3/2} -\frac{3}{2}A\rho\dot{r}\sqrt{l-r}
\end{equation}
with the dissipative constant $A$ being a function of the elastic and viscous material parameters \cite{brilliantov1996} and the other parameters as described before. The collision terminates at time $\tau$ when $\dot{r}(\tau)<0$ and $\ddot{r}(\tau)=0$, corresponding to purely repulsive interaction, see \cite{schwager2008}. The dissipative part, $F_n^{\text{dis}}$, was first motivated in \cite{kuwabara1987} and then rigorously derived in \cite{brilliantov1996} and \cite{morgado1997}, where only the approach in \cite{brilliantov1996} leads to an analytic expression of the material parameter $A$.

We apply the same scaling as in Sec. \ref{sec:Dimensionless} to obtain
\begin{equation}
\label{eq:newtonPolarScaledV}
\begin{split}
  \frac{\mathrm d \varphi}{\mathrm d
\tilde{t}}&=\frac{c_\varphi}{\tilde{r}^2}\\
 \frac{\mathrm d^2\tilde{r}}{\mathrm d\tilde{t}^2}&=
 \tilde{r}\left(\frac{\mathrm d \varphi}{\mathrm d \tilde{t}}\right)^2+
 \left(\tilde{l}-\tilde{r}\right)^{\nicefrac{3}{2}}-
 c_{\text{dis}}\sqrt{\tilde{l}-\tilde{r}}\frac{\mathrm d\tilde{r}}{\mathrm
d\tilde{t}}
\end{split}
\end{equation}
with the definitions and initial conditions given in Eqs. (\ref{eq:scalingDef}, \ref{eq:cPhiCDisDef}, \ref{eq:initCondScaled}) and additionally
\begin{equation}
\label{eq:CDisDef}
c_{\text{dis}}\equiv\gamma\sqrt{X}T\,;~~~~~\gamma\equiv\frac{3}{2}\frac{\rho A}{m_{\text{eff}}}\,.
\end{equation}

In contrast to the case of elastic spheres discussed in Section \ref{sec:Elastic}, for inelastic frictionless spheres we need three independent parameters to describe their collisions, $\tilde{l}$, $c_\varphi$ and $c_{\text{dis}}$.

\subsection{The Role of Inelasticity}
\label{sec:influenceOfInelasticity}

\begin{figure}[h!]
 \includegraphics[width=\figWidth]{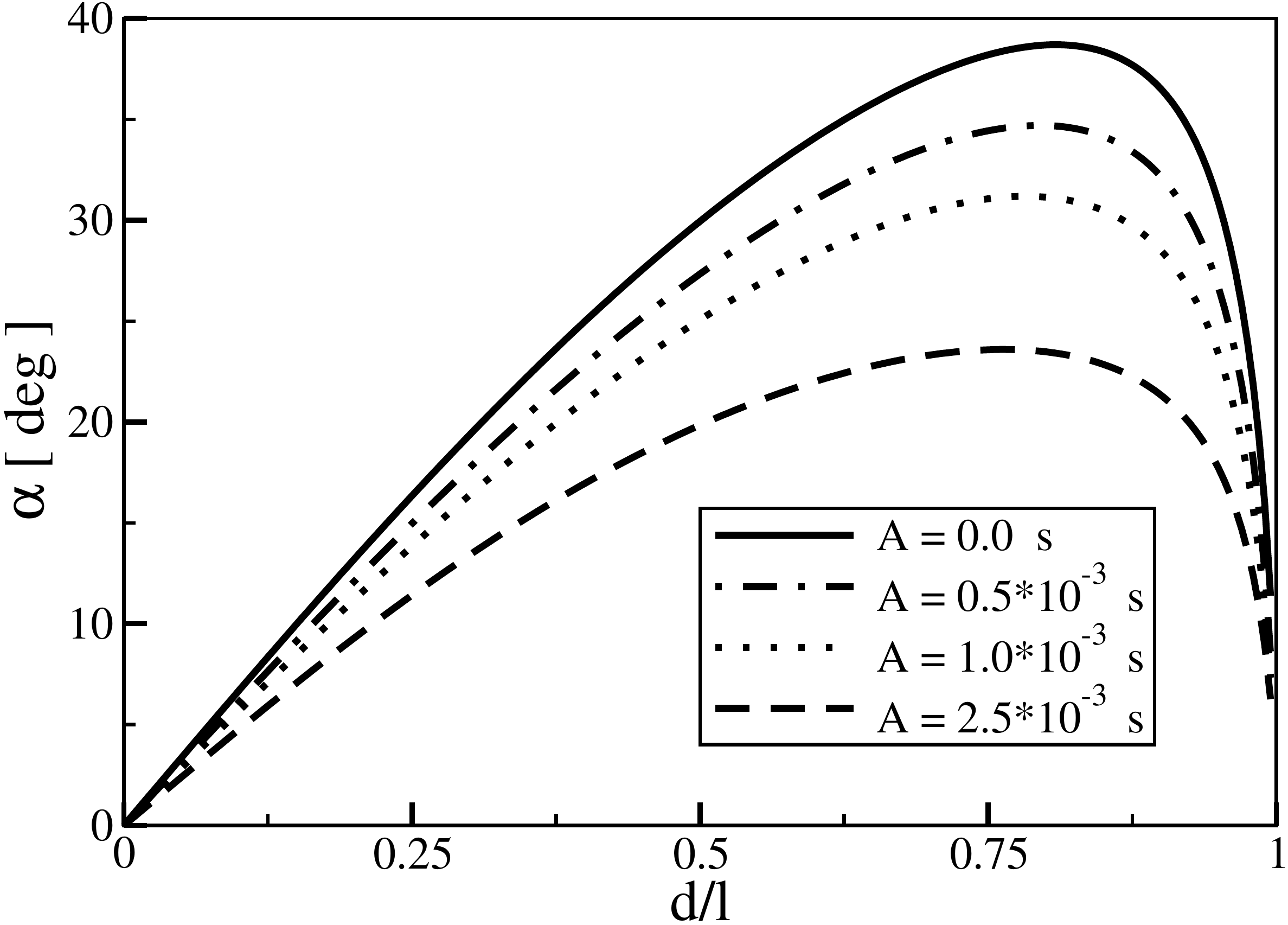}
\caption{Rotation angle $\alpha$ of the unit vector $\hat{e}_r$ as function of the eccentricity for various dissipative constants $A$. The elastic parameters are the same as for Fig. \ref{fig:alphaVsDOverD}.}
\label{fig:alphaVsDOverDdis}
\end{figure}

To study the dependence of the rotation angle $\alpha$ on the inelasticity, we repeat the computation shown in Sec. \ref{sec:Rotation} (same elastic parameters) for inelastic collisions where $A\ne 0$. Figure \ref{fig:alphaVsDOverDdis} shows the rotation of the unit vector $\hat{e}_r$ during inelastic collisions over the eccentricity $d/l$ (see \figRef{fig:excCollSetup}) for various dissipative constants $A$. 

\begin{figure}[h!]
 \includegraphics[width=\figWidth]{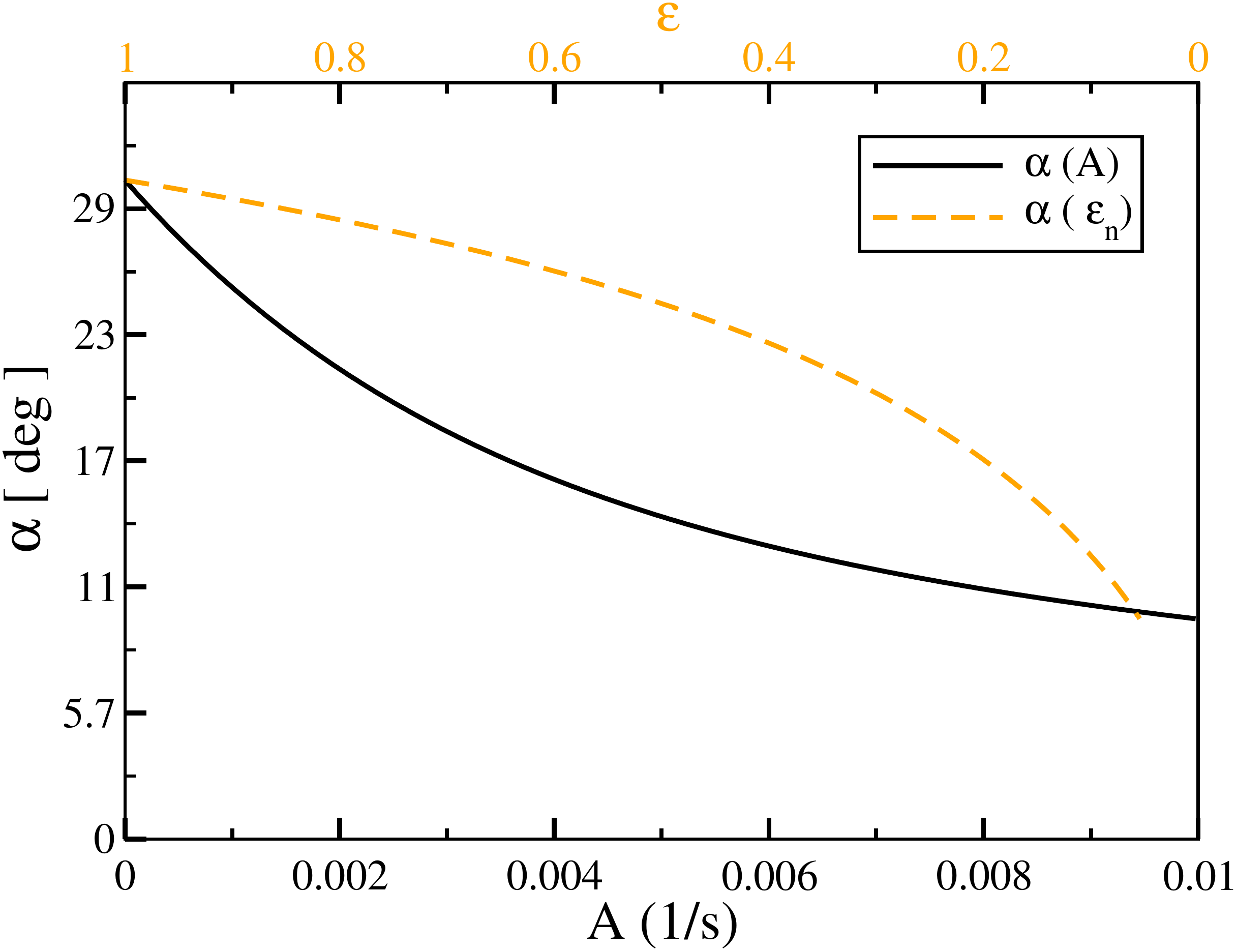}
\caption{Rotation angle $\alpha$ over the dissipative parameter $A$ (lower scale, full line) and over the coefficient of normal restitution $\varepsilon$ (upper scale, dashed line) for $d/l=0.5$.}
\label{fig:alphaVsA}
\end{figure}

In \figRef{fig:alphaVsA} we additionally fix $d/l=0.5$ to plot the rotation $\alpha$ as a function of the dissipative constant $A$. To provide a more vivid quantity for the inelasticity of the collision, we give $\alpha$ also as a function of the coefficient of restitution $\varepsilon$ corresponding to a central collision at the chosen impact velocity $v_n=\sqrt{3}/2 \cdot 20\:\text{m/s}\approx 17.3\:\text{m/s}$  \cite{Pade}.

Disregarding for a moment the centrifugal force, the dependence of the moment of inertia on $l(t)$ and the fact that the final deformation $l(\tau)$ depends on $A$ \cite{schwager2008}, the decreasing function $\alpha(A)$ or $\alpha(\varepsilon)$ may be understood essentially from the fact that the duration of the contact is a decreasing function of inelasticity, $d \tau(A)/d A <0$.  Thus, the smaller the coefficient of restitution the shorter lasts the contact and the smaller is the rotation angle during contact. This explanation is certainly oversimplified and serves only as a motivation to understand qualitatively the behavior of $\alpha(A)$.

As shown qualitatively in Fig. \ref{fig:alphaVsDOverDdis} and quantitatively in Fig. \ref{fig:alphaVsA}, for all eccentricities the rotation angle adopts its maximum for $A=0$, corresponding to elastic collisions, $\varepsilon=1$. 

\section{Conclusion}
\label{sec:conclusion}

For all real materials the collision of particles implies a rotation of the inter-particle unit vector $\hat{e}_r$ during the time of contact $\tau$ by a certain angle $\alpha$. This rotation is neglected in Kinetic Theory of granular systems as well as in event-driven Molecular Dynamics simulations relying both on the hard-sphere model of granular particles. Therefore, to justify the application of the hard-sphere model, one has to assure that the rotation angle is negligible for the given system parameters. In the present paper, we reduce the problem of oblique {\em elastic} collisions to two independent parameters, $\tilde{l}$ and $c_\varphi$, and compute the rotation angle $\alpha$ as a function of these parameters. The result is {\em universal}, that is, $\alpha$ is known for {\em any} combination of material parameters (Young modulus $Y$, Poisson ratio $\mu$, material density $\rho_m$) and system parameters (particle radii $R$, impact eccentricity $d/l$, and impact velocity $v$). 

For dissipative collisions characterized by the coefficient of restitution, $0<\varepsilon<1$, we show that the rotation angle is smaller than for the corresponding elastic case where all parameters are the same, except for $\varepsilon=1$. Therefore, to assess whether the rotation angle is small enough to justify the hard sphere approximation for a given system of dissipative particles, it is sufficient to consider the corresponding system of {\em elastic} particles discussed in Sec. \ref{sec:Elastic}.

For convenient use of our result we provide a universal lookup table and the corresponding access functions (see Online Resource \cite{supplement}). The angle of rotation for a given situation can be either obtained using the dimensionless variables, $\alpha(c_\varphi,\tilde{l})$, obtained from Eqs. \eqref{eq:scalingDef} and \eqref{eq:cPhiCDisDef} together with Eqs. \eqref{eq:rhodef} and \eqref{eq:XTdef} for the present material and system parameters, or by providing the physical parameters directly.

Concluding we consider our result as a tool to assess whether the Kinetic Theory description of a granular system on the basis of the Boltzmann equation and/or its simulation by means of highly efficient event-driven Molecular Dynamics is justified.

\begin{acknowledgement}
The authors gratefully acknowledge the support of the Cluster of Excellence 'Engineering of Advanced Materials' at the University of Erlangen-Nuremberg, which is funded by the German Research Foundation (DFG) within the framework of its 'Excellence Initiative'. 
\end{acknowledgement}

\bibliographystyle{spmpsci}      

\end{document}